\documentstyle{article}
\input math_macros
\font\tenbf=cmbx10
\font\tenrm=cmr10
\font\tenit=cmti10
\font\elevenbf=cmbx10 scaled\magstep 1
\font\elevenrm=cmr10 scaled\magstep 1
\font\elevenit=cmti10 scaled\magstep 1

\renewenvironment{thebibliography}[1]
 { \elevenrm
   \begin{list}{\arabic{enumi}.}
    {\usecounter{enumi} \setlength{\parsep}{0pt}
     \setlength{\itemsep}{3pt} \settowidth{\labelwidth}{#1.}
     \sloppy
    }}{\end{list}}

\begin{document}
\begin{center}{{\tenbf JET INCLUSIVE CROSS SECTIONS\\}
\vglue 5pt
\vglue 1.0cm
{\tenrm Vittorio Del Duca \\}
\baselineskip=13pt
{\tenit Stanford Linear Accelerator Center\\}
\baselineskip=12pt
{\tenit Stanford University, Stanford, California 94309\\}
\vglue 0.8cm}
\end{center}
{\rightskip=3pc
 \leftskip=3pc
 \tenrm\baselineskip=12pt
 \noindent
 Minijet production
 in jet inclusive cross sections at hadron colliders, with large
 rapidity intervals between the tagged jets, is evaluated by using the
 BFKL pomeron. We describe the jet inclusive cross section for an
 arbitrary number of tagged jets, and show that it behaves like a
 system of coupled pomerons.
\vglue 0.6cm}
\baselineskip=14pt
\elevenrm
At the present and next hadron colliders the semihard region,
characterized by an intermediate scale $m$, such that
$s >> m^2 >> \Lambda^2_{QCD}$, and large rapidity intervals,
of order $\ln(s/m^2)$, is expected to play an increasingly important
role, due to a copious production of minijets.
In this region scattering events are dominated by the
exchange of gluon ladders in the $t$ channel. In the leading
logarithmic approximation (LLA), which resums $log(s/t)$ terms,
the gluon ladders are described by the BFKL evolution equation$^1$.
The LLA is generated by the multi-Regge kinematics ,
i.e. by the strong ordering in rapidity of the produced minijets.

Then, according to the event we consider, we factorize
the cross section in such a way to have the large rapidity intervals
and the BFKL evolution either in the gluon structure functions$^2$
or in the short-distance cross section$^3$. In this contribution we
focus on the latter, namely we tag two or more jets at large
rapidity intervals$^4$, with the minimum transverse momentum
$m$ of the tagged jets as the factorization scale of the
short-distance cross section.

Mueller and Navelet$^3$ proposed to compute in this fashion the
2-jet inclusive cross
section. By keeping the light-cone momentum fractions $x_1$ and $x_2$
of the beam momenta fixed, such a cross section depends only on
$s$, i.e. on the interval in rapidity
$\eta = \ln(x_1x_2 s/m^2)$ between the tagged jets. These
lie at the extremes of the lego plot in rapidity and azimuthal angle,
with the rapidity interval between them filled with minijets.
Assuming that the hard scattering
is initiated by gluons, we can write the 2-jet inclusive cross section,
integrated over the transverse momenta of the tagged
jets, in a factorized form

\begin{equation}
\sigma(s,m^2,x_1,x_2) = x_1 G(x_1,m^2) x_2 G(x_2,m^2)
   \hat\sigma_{tot}(gg,\eta),
\end{equation}

with $G(x,m^2)$ the DGLAP gluon distribution and $\hat\sigma_{tot}
(gg,\eta_)$ the total gluon-gluon cross section, which involves only
the hard part of the scattering. At large
rapidity intervals, since the leading scattering always goes through
gluon exchange in the $t$ channel, we can
include initial-state quarks$^5$ by simply replacing $G(x,m^2)$ with
$G(x,m^2) + 4/9 \sum_f (Q_f(x,m^2) + \bar Q_f(x,m^2))$.

The total gluon-gluon cross section can then be computed from the
imaginary
part of the corresponding forward elastic scattering amplitude, with
color-singlet exchange in the $t$ channel, obtained by using the BFKL
pomeron$^1$. It can be written as

\begin{equation}
\hat\sigma_{tot}(gg,\eta) = {9\pi\alpha_s^2 \over 2m^2}
   f(\eta),
\end{equation}

where $f(\eta)$, the ratio of the total to the Born gluon-gluon cross
sections, is given in terms of the Fourier spectrum of the pomeron
off-shell amplitude convoluted with the Fourier structures of the
off-shell scattering amplitudes at the extremes of the lego plot, which
in this case are merely the $3$-gluon vertices,

\begin{equation}
f(\eta) = {1 \over 2\pi}
\int_{-\infty}^{\infty} d\nu {1 \over \nu^2+{1/4}} e^{\eta\omega(\nu)},
\end{equation}

and
\begin{equation}
\omega(\nu) = 6 {\alpha_s \over\pi} \bigl[ \psi(1) - Re\psi
   ({1\over 2} +i\nu) \bigr].
\end{equation}

The azimuthal angle between the transverse momenta of the tagged
jets has been integrated out.
At large values of $\eta$ the asymptotics of $f(\eta)$ is

\begin{equation}
f(\eta) = {e^{12\log{2}{\alpha_s\over\pi}\eta}\over
   \sqrt{{21\over 2}\zeta(3)\alpha_s\eta}},
\end{equation}

The exponential growth of $f(\eta)$ is due to the minijet production.

Since we are assuming no BFKL evolution to the exterior of the
left and rightmost tagged jets on the lego plot, Eq. (1) can be
generalized to determine the $(n+2)$-jet inclusive cross section from
the $n$-gluon inclusive one. For the $1$-gluon inclusive cross
section, for instance, fixing $\bar{\eta}$, $\eta_A$ and $\eta_B$ as the
rapidities of the gluon and the jets at the extremes of the lego
plot respectively, with $\eta_A > \bar{\eta} > \eta_B$ and
$\eta = \eta_A - \eta_B$, we have$^6$

\begin{eqnarray}
{d\sigma_1 \over d\bar{\eta} d\ln(k^2/m^2)} & = & {9\pi\alpha_s^2 \over 2m^2}
\frac{3\alpha_s}{4\pi^3} \int_{\infty}^{\infty}d\nu_A
\int_{\infty}^{\infty}d\nu_B e^{(\eta_A-\bar{\eta})\omega(\nu_A)}
e^{(\bar{\eta}-\eta_B)\omega(\nu_B)}e^{-i(\nu_A-\nu_B)\ln(k^2/m^2)}
\frac{1}{i(\nu_A-\nu_B)} \nonumber \\ & & \left[ \frac{1}{1/2-i\nu_A}
\frac{\Gamma(1/2-i\nu_A)}{\Gamma(1/2+i\nu_A)}
\frac{\Gamma[1+i(\nu_A-\nu_B)]}{\Gamma[1-i(\nu_A-\nu_B)]} \frac
{\Gamma(1/2+i\nu_B)}{\Gamma(1/2-i\nu_B)}\frac{1}{1/2+i\nu_B} \right],
\end{eqnarray}

with $Im(\nu_A-\nu_B) < 0$ in order not to have unphysical
ultraviolet divergences, and $Im\nu_A > -1/2$ and $Im\nu_B < 1/2$.
Thus the 1-gluon inclusive cross section can be described in terms
of two coupled pomerons, which resonate when their Fourier frequencies
are about the same. At large intervals in rapidity, it is possible to
perform a saddle-point evaluation of Eq.(6)

\begin{equation}
{d\sigma_1 \over d\bar{\eta} d\ln(k^2/m^2)} = \hat\sigma_{tot}(gg,\eta)
\frac{3\alpha_s}{2\pi^{5/2}} \sqrt{1-{1\over a}} exp\left[ -{1\over a-1}
+ 1/2 \right],
\end{equation}
with
\begin{equation}
a = \sqrt{1+{336\zeta(3)\alpha_s(\eta_A-\bar{\eta})(\bar{\eta}-\eta_B)
    \over \pi\eta \ln^2(k^2/m^2)}}.
\end{equation}

Eq.(7) shows that at fixed gluon rapidities and
large transverse momenta the 1-gluon inclusive cross section falls
off faster than any power of the momentum. However, the requirement
that the saddle points, which lie on the imaginary axis, don't hit the
poles makes the saddle-point evaluation feasible only in a region
central in rapidity and shrinking as the gluon transverse momentum grows.
This can be avoided by evaluating Eq.(6) numerically.

Eq.(6) is straightforwardly generalizable to the $n$-gluon
inclusive cross section, with a strong ordering of the Fourier
frequencies of the exchanged pomerons on the imaginary axis.
The region where a saddle-point
evaluation is feasible becomes quickly negligible as the number of
tagged jets grows and a numerical evaluation of Eq.(6) is then
compelling.

What said above can be generalized to more elaborate events, like
heavy quark or Higgs boson production, via gluon-gluon fusion.
For scattering events with additional intermediate scales, like in
the inclusive production of a Higgs boson and a gluon jet$^7$ at the
extremes of the lego plot, the BFKL pomeron may also resum the
collinear enhancements which arise in the short-distance cross section
because of the higher intermediate scale$^8$, in this case the
mass of the top quark in the gluon-gluon-Higgs form factor.
Then following the
procedure outlined above it is possible to consider also the production of
either the Higgs boson or more gluon jets in the middle of the lego plot.

\vglue 0.6cm

Work supported by the Department of Energy, contract DE-AC03-76SF00515.
The author wishes to thank Bj Bjorken, John Collins, Keith Ellis, Lev
Lipatov, Al Mueller, Michael Peskin, Carl Schmidt, George Sterman and
Wai Tang for many interesting and stimulating discussions.

\refend

\begin{thebibliography}{9}
\bibitem{BFKL}L.N. Lipatov, {\elevenit Sov. J. Nucl. Phys.}
{\elevenbf 23} (1976) 338;

E.A. Kuraev, L.N. Lipatov and V.S. Fadin, {\elevenit Sov. Phys. JETP}
{\elevenbf 44} (1976) 443;

E.A. Kuraev, L.N. Lipatov and V.S. Fadin, {\elevenit Sov. Phys. JETP}
{\elevenbf 45} (1977) 199;

Ya. Ya. Balitskii and L.N. Lipatov, {\elevenit Sov. J. Nucl. Phys.}
{\elevenbf 28} (1978) 822;

L.N. Lipatov, {\elevenit Sov. Phys. JETP} {\elevenbf 63} (1986) 904.
\bibitem{er} J.C. Collins and R.K. Ellis, {\elevenit Nucl. Phys.}
{\elevenbf B360} (1991) 3;

S. Catani, M. Ciafaloni and F. Hautmann, {\elevenit
Nucl. Phys.} {\elevenbf B366} (1991) 135.
\bibitem{MN} A.H. Mueller and H. Navelet, {\elevenit Nucl. Phys.}
{\elevenbf B282} (1987) 727.
\bibitem{bj}J.D. Bjorken, {\elevenit Int. J. Mod. Phys.}
{\elevenbf A7} (1992) 4189.
\bibitem{CM}B.L. Combridge and C.J. Maxwell, {\elevenit Nucl. Phys.}
{\elevenbf B239} (1984) 429.
\bibitem{VM} V. Del Duca, M.E. Peskin and W.-K. Tang, in progress.

\bibitem{daw}
%
%
%
%
S. Dawson and R.P. Kaufmann, {\elevenit Phys. Rev. Lett.} {\elevenbf
68} (1992) 2273.
\bibitem{DS}V. Del Duca and C. Schmidt, in progress.
\end{thebibliography}
\end{document}